\documentclass[aps,superscriptaddress]{revtex4-2}
\usepackage{graphicx,color}
\usepackage{psfrag}

\def\beq{\begin{equation}}
\def\eeq{\end{equation}}

\begin{document}

\title{Entanglement of bosonic systems under monitored evolution}

\author{Quancheng Liu}
\affiliation{School of Physics, State Key Laboratory of Crystal Materials, Shandong University, Jinan, 250100, China}
\affiliation{Department of Physics, Institute of Nanotechnology and Advanced Materials, Bar-Ilan University, Ramat-Gan 52900, Israel}

\author{Klaus Ziegler}
\affiliation{Institut f\"ur Physik, Universit\"at Augsburg, D-86135 Augsburg, Germany}

\begin{abstract}
The evolution of 
non-interacting bosons in the presence of repeated projective measurements is studied.
Following the established approach, this monitored evolution is characterized by the first detected return
and the first detected transition probabilities. We show that these quantities are directly related to the
entanglement entropy and to the entanglement spectrum of a bipartite system. Calculations with specific values
for the number of bosons, the number of measurements and the time step between measurements reveal a 
sensitive and often strongly fluctuating entanglement entropy. 
In particular, we demonstrate that in the vicinity of special values for the time steps  the evolution of the entanglement 
entropy is either stationary or performs dynamical switching between two or more stationary values.
In the entanglement spectrum, on the other hand, this complex behavior can be associated with level
crossings, indicating that the dominant quantum states and their entanglement respond strongly to
a change of the system parameters. We discuss briefly the role of time averaging to remove the fluctuations
of the entanglement entropy.  
\end{abstract}


\maketitle

\section{Introduction}

Repeated measurement on a quantum system has been used to determine the first detected return (FDR) to the initial
state or the first detected transition (FDT) to a state that is different from the initial state. 
The idea is to prepare the 
quantum system in an initial state $|\psi_0\rangle$, let it evolve unitarily to the state
$e^{-iH\tau}|\psi_0\rangle$ and perform a projective measurement with the projector $\Pi={\bf 1}-|\psi\rangle\langle\psi|$,
where ${\bf 1}$ is the identity operator and $|\psi\rangle$ a state that defines the measurement. This operation yields
the state $|\psi_1'\rangle=\Pi e^{-iH\tau}|\psi_0\rangle$, which is either orthogonal to $|\psi\rangle$ or it vanishes
when $e^{-iH\tau}|\psi_0\rangle=e^{i\varphi}|\psi\rangle$ with some phase $\varphi$. A further unitary evolution 
for the time $\tau$ yields $|\psi_1\rangle=e^{-iH\tau}\Pi e^{-iH\tau}|\psi_0\rangle$ and 
$\phi_1=\langle\psi|\psi_1\rangle$. If $\phi_1\ne 0$ the system was not in the state 
$e^{i\varphi}|\psi\rangle$ when the projection was applied. This means that our measurement to detect $|\psi\rangle$
was not successful. In this case we apply another projection to $|\psi_1\rangle$, followed by
a unitary evolution to get $|\psi_2\rangle=e^{-iH\tau}\Pi|\psi_1\rangle$ and $\phi_2=\langle\psi|\psi_2\rangle$.
Again, if $\phi_2\ne0$ the system was not detected in the state $e^{i\varphi}|\psi\rangle$. These
steps can be repeated $m$ times until $\phi_k=0$ for all $k\ge m$. In other words, 
if the measurement is unsuccessful by not detecting the state $|\psi\rangle$, the experiment continues by 
another measurement, followed by the evolution for time step $\tau$. 
This protocol was discussed in Ref.~\cite{Gruenbaum2013} and has been applied to single-particle states to detect 
the particle location on a 
graph~\cite{Dhar_2015,dhar15,Friedman_2016,lahiri19,PhysRevResearch.1.033086,PhysRevResearch.2.033113,
PhysRevResearch.5.023141}.

For the evolution of a system with more than one particle the entanglement of the quantum state is a fundamental property.
It can be characterized, for instance, by the R\'enyi entropy that measures the quantum 
correlations between two subsystems under a spatial
bipartition~\cite{PhysRevD.34.373,PhysRevLett.71.666,PhysRevA.66.042327,RevModPhys.82.277,PhysRevLett.127.040603}.
Probing the entanglement entropy (EE) has become an important and popular concept to study measurement-induced 
entanglement transitions,
to characterize the many-body evolution and many-body 
localization~\cite{
RevModPhys.80.517,RevModPhys.91.021001,PhysRevLett.90.227902,
calabrese2004,peschel2009,calabrese2009,PhysRevX.5.041047,
doi:10.1146/annurev-conmatphys-031214-014726,
PhysRevX.8.021062,PhysRevX.8.041019,
PhysRevB.100.134306,
PhysRevX.9.031009,
PhysRevB.104.155111,
PhysRevA.107.012413,PhysRevC.107.L061602,PhysRevC.107.044318}
and to classify the topology of quantum systems.

In this paper we will study the R\'enyi entropy in a system of non-interacting
bosons, which is subject to periodically repeated projective measurements. This enables us to study the FDR/FDT 
probabilities as well as the EE and entanglement spectrum (ES), and to investigate the relation between 
both quantities.

The focus of this paper is on physical aspects of the monitored evolution that can also be applied to quantum 
computing~\cite{PhysRevResearch.5.033089}. Although quantum computing is typically based on qubits (i.e., spin states),
potentially bosonic systems could also be used~\cite{10.21468/SciPostPhys.12.1.009}.
A promising example are photonic states. 
In particular, we consider $N$ non-interacting bosons, distributed  in two wells which are coupled by tunneling.
This can be experimentally realized as a pair of photonic cavities that are coupled by an optical 
fiber~\cite{PhysRevA.88.013834,Haroche2020,PhysRevLett.106.240505,PhysRevLett.109.020502,Cao:20}.
The underlying Hilbert space is $N+1$-dimensional and enables us to study the scaling behavior of the 
entanglement with $N$.

The structure of the paper is that in Sect. \ref{sect:concept} we provide a general theory part with the 
definitions of the FDR/FDT 
probabilities and their relations to the reduced density matrix, the EE and the ES.
More details for the calculation of the FDR/FDT probabilities are provided in Sect. \ref{sect:fdr/fdt}. This includes an approach
that connects the monitored evolution due to repeated measurements to the unitary evolution. 
The reader, who is not interested in the theoretical concepts but more in the results of the monitored evolution, can skip 
this section.
Then in Sect. \ref{sect:model} our approach to the monitored evolution is applied to the tunneling of $N$ 
non-interacting bosons in a double well. For this specific model the eigenvalues and spectral weights are calculated.
Finally, in Sect. \ref{sect:discussion} specific examples in terms of the parameters of the model are presented for the EE 
and the ES.

\section{General concept of projected measurements}
\label{sect:concept}

The quantum system is characterized by the density operator that reads in the presence of repeated projective measurements
\beq
\label{density_operator00}
\rho^m(\tau)
=\frac{1}{{\cal N}}
e^{-iH\tau}(\Pi e^{-iH\tau})^{m-1}|\psi_0\rangle\langle\psi_0|(e^{iH\tau}\Pi)^{m-1}e^{iH\tau}
\eeq
with the normalization 
${\cal N}=Tr[e^{-iH\tau}(\Pi e^{-iH\tau})^{m-1}|\psi_0\rangle\langle\psi_0|(e^{iH\tau}\Pi)^{m-1}e^{iH\tau}]$.
This density operator describes a quantum walk~\cite{PhysRevA.48.1687}, where after each time steps $\tau$
a projective measurement $\Pi$ 
is applied. The latter prevents the visit of the Hilbert space after a time step $\tau$ that is orthogonal to
the $\Pi$-projected Hilbert space. 
Although in general the projector $\Pi$ is independent of the initial state $|\psi_0\rangle$ and can be chosen
freely, in this paper we will focus on the case discussed in the Introduction, where $\Pi$ projects onto the Hilbert 
space which is orthogonal to a given state $|\psi\rangle$ as $\Pi={\bf 1}-|\psi\rangle\langle\psi|$.

For the following discussion we start from a product space ${\cal H}_1\otimes{\cal H}_2$ with $n_1$ bosons in the
left reservoir and $n_2$ bosons in the right reservoir.
Then we assume that the Hamiltonian obeys particle conservation $n_1+n_2=N$, which implies that it acts inside the Hilbert 
space that is spanned by the basis $\{|n,N-n\rangle\}_{0\le n\le N}$.
The basis states can also contain additional quantum numbers, which are not relevant for the general discussion.
In Sect. \ref{sect:model} we will consider the special case in which the Hilbert space represents a 
double well, where $n$ bosons are in the left and $N-n$ bosons are in the right well. Then the basis is constructed
from Fock states as $|n,N-n\rangle\equiv|n\rangle|N-n\rangle$ without additional quantum numbers.

Returning to the general case,
in the basis $\{|n,N-n\rangle\}_{0\le n\le N}$ the $(N+1)\times (N+1)$ density matrix reads 
$\rho^m_{n,n-N;n',N-n'}=\langle n,N-n|\rho^m(\tau)|n',N-n'\rangle$ with $n,n'=0,\ldots ,N$. 
After summing over all basis states of ${\cal H}_2$ the reduced density matrix ${\hat\rho}$ 
becomes an $(N+1)\times(N+1)$ diagonal matrix with elements
\beq
\label{reducedDM}
{\hat\rho}^m_{nn}
=\sum_{n'=0}^N\langle n,n'|\rho^m(\tau)|n,n'\rangle
=\langle n,N-n|\rho^m(\tau)|n,N-n\rangle
\]
\[
=\frac{1}{\cal N}\langle n,N-n|e^{-iH\tau}(\Pi e^{-iH\tau})^{m-1}|\psi_0\rangle\langle\psi_0|(e^{iH\tau}\Pi )^{m-1}e^{iH\tau}
|n,N-n\rangle
.
\eeq
With the projector $P_n:=|n,N-n\rangle\langle n,N-n|$ the density matrix elements can also be written as a trace expression
\beq
\label{trace_expression}
{\hat\rho}^m_{nn}
=\frac{1}{\cal N}Tr \left[P_ne^{-iH\tau}(\Pi e^{-iH\tau})^{m-1}P_0(e^{iH\tau}\Pi)^{m-1}e^{iH\tau}\right]
,
\eeq
where we have assumed $|\psi_0\rangle=|0,N\rangle$ for the initial state. 
One should note that the right-hand side of Eq. (\ref{reducedDM}) with $\Pi$ replaced by $\Pi_n={\bf 1}-P_n$
is either the first detected return (FDR) probability (for $n=0$) or the first detected transition (FDT) probability (for $n>0$).
Therefore, known results of  the FDR/FDT 
probabilities~\cite{Grunbaum2014,Friedman2017,PhysRevResearch.1.033086} can be directly used for the reduced 
density matrix through the relation
\beq
\label{FDR/FDT_relation}
{\hat\rho}^m_{nn}=\frac{|\phi_{m;n0}|^2}{\sum_{n=0}^N|\phi_{m;n0}|^2}
\eeq
with the FDT amplitude for $|0,N\rangle\to |n,N-n\rangle$ ($n\ne 0$) after $m$ measurements
\beq
\label{transition_amp1}
 \phi_{m;n0}:=\langle n,N-n|e^{-iH\tau}(\Pi_n e^{-iH\tau})^{m-1}|0,N\rangle
\eeq
and the corresponding FDR amplitude $\phi_{m;00}$ for $|0,N\rangle\to |0,N\rangle$.
It is crucial to note that $\sum_{n=0}^N|\phi_{m;n0}|^2\le 1$ for $m>1$ due to the projection $\Pi$.
Some known results for the FDR/FDT amplitudes are summarized in Sect. \ref{sect:fdr/fdt}.

With this expression for ${\hat\rho}^m_{nn}$ we can introduce the  
R\'enyi entropy~\cite{PhysRevX.8.041019} as a quantitative measure for the entanglement of the two Hilbert spaces
${\cal H}_1$ and ${\cal H}_2$
\begin{equation}
\label{ent_entropy}
	{\cal S}_\alpha (\tau,N,m)
=\frac{1}{1-\alpha}\log_2 {\rm Tr}[{({\bar\rho}^m)^\alpha(\tau)}] 
.
\end{equation}
In general, $\alpha$ is a free parameter and typical values used are $\alpha=2,3$~\cite{PhysRevX.8.041019}. 
For the subsequent calculations we set $\alpha=2$, i.e., we will calculate ${\cal S}_2(\tau,N,m)$ as the EE.

While the EE reveals a global measure for the monitored evolution, the ES~\cite{PhysRevLett.101.010504} 
provides a local measure of the evolution for the transition between individual states $|0,N\rangle\to|n,N-n\rangle$.
In other words, it reveals the contribution of individual states to the entangled
state of the evolving quantum system. It is defined as the logarithm of the reduced density matrix eigenvalues.
In the present case the reduced density matrix is already diagonal, such that
\beq
\label{espectrum}
\xi_{m;n} = -\log ({\bar\rho}^m_{nn})
=-2\log(|\phi_{m;n0}|)+\log\left(\sum_{n=0}^N|\phi_{m;n0}|^2\right)
.
\eeq
The smallest value of $\xi_{m;n}$ corresponds with the dominant transition $|0,N\rangle\to|n,N-n\rangle$
after $m$ measurements. A crossing of the lowest levels upon changing of the time step $\tau$ or $m$ is
reminiscent of a phase transition in classical statistical systems due to a crossing of the ground state energies.

The FDR/FDT amplitude in Eq. (\ref{transition_amp1}) is written in the Fock basis. It is convenient to express the
evolution operator in the eigenbasis $\{|E_k\rangle\}$ of the Hamiltonian as
\beq
\phi_{m;n0}=\sum_{\{k_j=0\}}^N
\langle\psi_n|E_{k_1}\rangle e^{-iE_{k_1}\tau}\langle E_{k_1}|\Pi_n|E_{k_2}\rangle e^{-iE_{k_2}\tau}\cdots
\langle E_{k_{m-1}}|\Pi_n|E_{k_{m}}\rangle e^{-iE_{k_{m}}\tau}\langle E_{k_{m}}|\psi_0\rangle
.
\eeq
For the projector $\Pi_n={\bf 1}-|n,N-n\rangle\langle n,N-n|$ a typical matrix element then reads
\beq
\langle E_k|({\bf 1}-|n,N-n\rangle\langle n,N-n|)|E_{k'}\rangle
=\delta_{kk'}-q^*_{n,k}q_{n,k'}=:({\bf 1}-Q_n^*EQ_n)_{kk'}
\eeq
with $q_{n,k}=\langle n,N-n|E_k\rangle$ and $q^*_{n,k}=\langle E_k|n,N-n\rangle$.
$E$ is the $(N+1)\times(N+1)$ matrix, whose matrix elements are 1. Moreover, $Q_n$ is a diagonal matrix,
consisting of the elements $\{q_{n,k}\}$. Then the FDR/FDT amplitude of Eq. (\ref{transition_amp1}) can be written as
\beq
\label{transition_amp2}
\phi_{m;n0}
=\sum_{k,k'=0}^Nq_{n,k}D_k[({\bf 1}-Q_n^*EQ_n)D]^{m-1}_{kk'}q^*_{0,k'}
\eeq
with $D_k=e^{-iE_k\tau}$. 
It should be noticed that $(Q_n^*)^{-1}(D-Q_n^* EQ_nD)^{m-1}Q_n^*$ is a function of $Q_nQ_n^*$
(cf. Ref. \cite{Ziegler_2021}): Since $Q_n$, $D$ are diagonal matrices, they commute and we get the relation
\beq
\label{relation3}
(D-Q_n^* EDQ_n)^{m-1}=Q_n^*(D-EDQ_nQ_n^*)^{m-1}(Q_n^*)^{-1}
,
\eeq
which is proved by complete induction. Moreover, we can write
$(D-Q^*_n EDQ_n)^{m-1}=D^{-1/2}T_n^{m-1}D^{1/2}$ with $T_n:=D^{1/2}({\bf 1}-Q^*_nEQ_n)D^{1/2}$ and for Eq. (\ref{transition_amp2})
\beq
\label{transition_amp3}
\phi_{m;n0}
=Tr (D^{1/2}T_n^{m-1}D^{1/2}Q^*_0EQ_n)
.
\eeq
Thus, the matrix $T_n$ represents the monitored evolution under repeated measurements,
which is the analogue of the unitary evolution matrix $U=\exp(-iH\tau)$ in the Fock basis.
Its largest eigenvalues control the large $m$ (i.e., stationary) behavior of $\phi_{m;n0}$, while the smaller eigenvalues decay
quickly. Therefore, the task is to identify those largest eigenvalues for the given parameters of the model. 

The matrix $T_n$ has some important properties that are useful for the calculation of $\phi_{m;n0}$. First,  
since $\sum_k q_{n,k}q^*_{n',k}=\delta_{nn'}$, there is a
complete set of right/left eigenvectors of the Hermitean matrix ${\bf 1}-Q^*_nEQ_n$ with an eigenvalue 0 and 
$N$ eigenvalues 1 due to
\beq
\label{eigenvectors}
\sum_{k'}(\delta_{kk'}-q^*_{n,k}q_{n,k'})q^*_{n',k'}=(1-\delta_{nn'})q^*_{n',k}
\ {\rm and}\ \
\sum_{k}q_{n',k}(\delta_{kk'}-q^*_{n,k}q_{n,k'})=(1-\delta_{nn'})q_{n',k'}
.
\eeq
Defining the vector ${\bf q}_n:=(q_{n,0},q_{n,1},\ldots,q_{n,N})^T$, we get
$Q^*_nEQ_n{\bf q}^*_{n'}={\bf q}_n\delta_{nn'}$, which implies for the vector 
${\bf x}:=\sum_{n'}a_{n'}D^{-1/2}{\bf q}^*_{n'}$
\beq
T_n{\bf x}=\cases{
D{\bf x} & for $a_n=0$ \cr
0 & for $a_{n'}=0$ ($n'\ne n$) \cr
}
.
\eeq
It is possible that $T_n{\bf x}=D{\bf x}$ reduces to $T_n{\bf x}=e^{i\varphi}{\bf x}$ when
we have $D{\bf x}=e^{i\varphi}{\bf x}$, which can happen in the presence of degenerate $E_j\tau\ ({\rm mod}\ 2\pi)$ or
when some components of the vectors ${\bf q}_{n'}$ vanish. In this case ${\bf x}$ is an eigenvector of 
$T_n$ whose eigenvalue lies on the unit circle of the complex plane. This means that ${\bf x}$ does not decay but 
accumulates only a phase. This eigenvector does not contribute to the FDR/FDT amplitude though, since ${\bf x}$ is 
orthogonal to ${\bf q}_n$ and, therefore, $EQ_n{\bf x}=0$.

Besides these special vectors, the eigenvalues of $T_n$ can be quite complex in general. 
On the other hand, even a two-level system (i.e., $N=1$, $n=0,1$) with energies $E_{0,1}$ is already instructive. 
In this case $T_n$ has two eigenvalues:
\beq
\lambda_0=0
\ ,\ \ 
\lambda_1=(1-|q_{n,0}|^2)e^{-2iE_0\tau}+|q_{n,0}|^2e^{-2iE_1\tau}
,
\eeq
where $|q_{n,1}|^2=1-|q_{n,0}|^2$ and $|\lambda_1|^2=1-2(1-|q_{n,0}|^2)|q_{n,0}|^2(1-\cos[2(E_1-E_0)\tau])$.
Thus, the eigenvalues of $T_n$ are on the unit disk, one at the center and the other one only for special values on the
unit circle, namely for $|q_{n,0}|=0,1$ and/or for $ (E_1-E_0)\tau=0\ ({\rm mod}\ \pi)$. This means that, except for the
special values with $|\lambda_1|=1$, $T_n^{m-1}$ decays exponentially fast. We will see subsequently that this
type of behavior exits also for larger systems. In particular, we will study a system of $N$ non-interacting bosons.

\subsection{FDR/FDT amplitudes}
\label{sect:fdr/fdt}

\begin{figure}[t]
\begin{center}
\includegraphics[width=0.5\columnwidth]{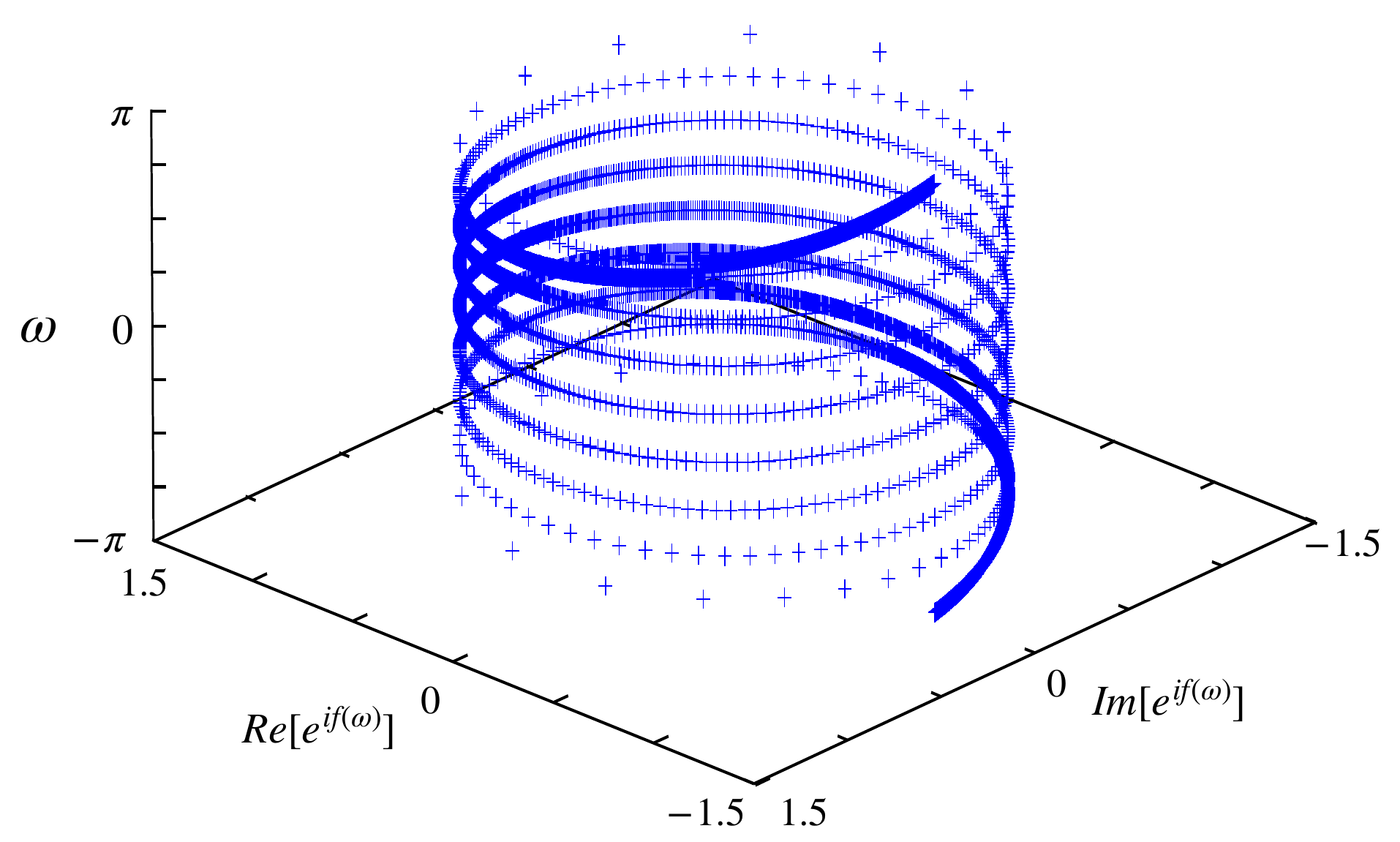}
\caption{${\hat \phi}_{r}(e^{i\omega})=e^{if(\omega)}$ for $-\pi\le\omega <\pi$ (vertical axis) 
with winding number $9$ for 8 non-interacting bosons.
The density of points is inverse to the sensitivity of the phase $f(\omega)$ to a change of $\omega$.}
\label{fig:phase2}
\end{center}
\end{figure}

Next we discuss the connection between the unitary and the monitored evolution by a linear relation.
We consider only the FDR/FDT amplitude, since this is the building block for the other physical quantities, according
to our discussion in the previous section.
The `first detected passage time problem', as discussed in Refs.~\cite{Gruenbaum2013,Friedman_2016}
for a single particle on a tight-binding graph, can be directly generalized to the evolution in a general Hilbert space.
The unitary evolution of the transition $|\psi_0\rangle\to|\psi_n\rangle$ for the time $\tau$ provides the amplitudes
\beq
\label{matrix_elements}
v_m:=\phi_{1;n0}(m\tau)
=\langle\psi_n|e^{-iHm\tau}|\psi_0\rangle
\ ,\ \ 
u_m:=\phi_{1;00}(m\tau)
=\langle\psi_0|e^{-iHm\tau}|\psi_0\rangle
.
\eeq
There exists a  mapping from the unitary amplitudes in Eq. (\ref{matrix_elements}) to the FDR/FDT amplitudes
$\phi_{m;n0}$
as (cf. App. \ref{app:expansion})
\beq
\label{final_eq}
{\vec \phi}=({\bf 1}+\Gamma)^{-1}{\vec v}
\eeq
with the $m$-component vectors ${\vec\phi}=(\phi_{1;n0},\phi_{2;n0},...,\phi_{m;n0})$, 
${\vec v}=(v_1,v_2,...,v_m)$ and with the triangular matrix $\Gamma$ whose elements are
\beq
\label{gamma_matrix}
\Gamma_{ij}=\cases{
u_{i-j} & $1\le i-j\le m-1$ \cr
0 & otherwise \cr
}
.
\eeq
In other words, the FDR/FDT amplitudes can be recursively constructed from the unitary amplitudes in
Eq. (\ref{final_eq}). This is solved by a discrete Fourier transformation with
\beq
\label{solution1}
\sum_{m\ge 1}z^m\phi_{m;n0}
={\hat\phi}(z)=\frac{{\hat v}(z)}{1+{\hat u}(z)}
\eeq
for $z$ inside the complex unit disk (i.e., $|z|<1$), and with the Fourier transformed unitary amplitudes 
of Eq. (\ref{matrix_elements}):
\beq
\label{spectr_rep3a}
{\hat u}(z)=z\sum_{j=0}^N \frac{|\langle\Psi_0|E_j\rangle|^2}{e^{iE_j\tau}-z}
\ ,\ \ \ 
{\hat v}(z)=z\sum_{j=0}^N \frac{\langle \Psi|E_j\rangle\langle E_j|\Psi_0\rangle}{
e^{iE_j\tau}-z}
.
\eeq 
The advantage of using a continuous function ${\hat\phi}(z)$ rather than a discrete function $\phi_{m;n0}$ is that analytic
tools, such as integration or perturbation theory, can be employed to exploit its properties.

Once the function ${\hat\phi}(z)$ is known,  $\phi_{m;n0}$ can be retrieved as the residue of the Cauchy integral:
Since ${\hat\phi}(z)$ does not have poles inside the unit disk due to 
$1+{\hat u}=\sum_j |\langle\Psi_0|E_j\rangle|^2/[1-z\exp(-iE_j\tau)]$, the FDR/FDT amplitude reads
\beq
\label{solution2}
\phi_{m;n0}=\frac{1}{2\pi i}\int_{C}z^{-m-1}\frac{{\hat v}}{1+{\hat u}}dz
\eeq
with a contour $C$ around $z=0$ smaller than the unit circle in order to avoid the
poles of ${\hat v}/(1+{\hat u})$. The function ${\hat\phi}_u={\hat u}/(1+{\hat u})$ is a uni-modular
function of the form ${\hat \phi}_{u}(e^{i\omega})=e^{if(\omega)}$ with a characteristic winding number that
is equal to the dimensionality of the underlying Hilbert space in the absence of degeneracies for 
$E_j\tau\ ({\rm mod}\ 2\pi)$~\cite{PhysRevResearch.1.033086}. The
example of $N=8$ non-interacting bosons is visualized in Fig. \ref{fig:phase2}.

\section{Physical model: non-interacting bosons in a double well}
\label{sect:model}

\begin{figure}[t]
\includegraphics[width=1\columnwidth]{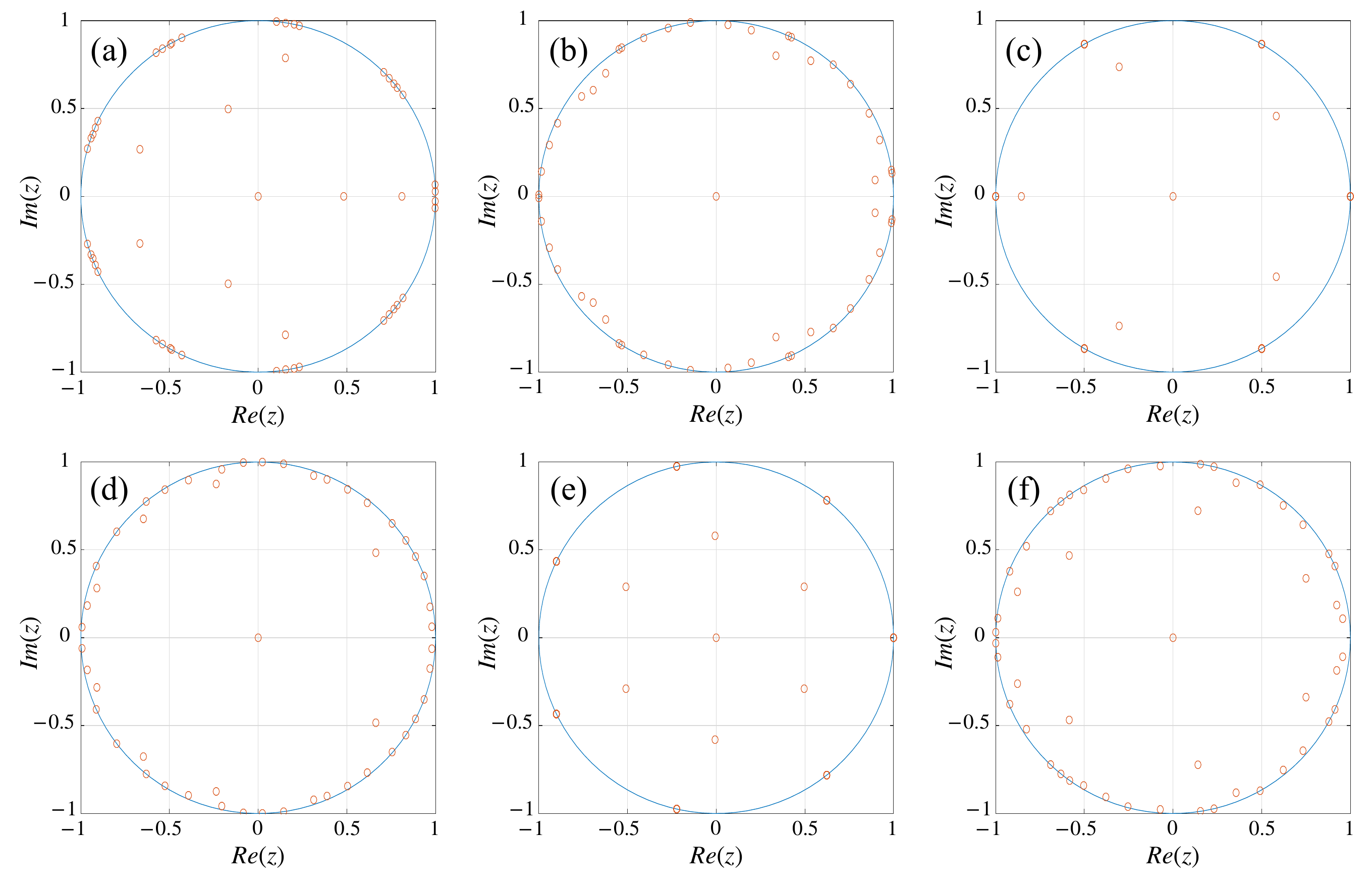}
\caption{
51 eigenvalues $z$ of $T$ for $N=50$ bosons with (a) $J\tau=0.7\hbar$, (b) $J\tau=0.5\hbar$,
(c) $J\tau=\pi\hbar/6$, (d) $J\tau=(\pi/6+10^{-2})\hbar$, (e) $J\tau=\pi\hbar/7$ 
and( f) $J\tau=(\pi/7+10^{-2})\hbar$. Very small changes in $J\tau$ have a strong effect on the eigenvalues
and their degeneracy. This can be seen by comparing (c) and (d) or (e) and (f). 
}
\label{fig:ee0}
\end{figure}

Within the single mode approximation~\cite{PhysRevA.55.4318}, 
the double well with $N$ bosons can be described as a two-site Bose model,
\begin{equation}
\label{ham00}
	H=-J(a_l^\dagger a_r + a_r^\dagger a_l)
\end{equation}
where $a^\dagger_{L,R}$ ($a_{L,R}$) are the bosonic creation (annihilation) operators in the left/right potential wells, and
$n_{l,r}=a_{l,r}^\dagger a_{l,r}$ are the corresponding number operators. $J$ is the tunneling rate of bosons between 
the two wells.
Using Fock states $|n,N-n\rangle \equiv |n\rangle|N-n\rangle$ ($n=0,\cdots,N$)
as a basis of the Hilbert space, the corresponding Hamiltonian matrix has a tridiagonal structure with 
\beq
\label{hamiltonian}
H_{n,n'}=\langle n,N-n|H|n',N-n'\rangle
= 
-J\sqrt{n(N+1-n)} \delta_{n,n'-1}  -J\sqrt{n'(N+1-n')} \delta_{n,n'+1} 
.
\eeq
These matrix elements represent an $(N+1)$-site tight-binding chain with broken translational invariance and
nearest-neighbor tunneling rates $-J\sqrt{n(N+1-n)}$.
The `sites' $n$ and $n'$ are connected by hopping of a single particle.

For non-interacting bosons we can calculate ${\hat\rho}_{m;nn}$ explicitly, since  the energy levels and
the spectral weight factors $\langle n,N-n|E_k\rangle$ are known. 
This system can be realized for photons at a beam splitter~\cite{PhysRevLett.57.13,haroche06} 
or in two harmonic cavities, which are connected through an optical 
fiber~\cite{PhysRevA.88.013834,Haroche2020,PhysRevLett.106.240505,PhysRevLett.109.020502,Cao:20}. 
Then the tunneling Hamiltonian $H=-J(a^\dagger_la^{}_r+a^\dagger_ra^{}_l)$
for $N$ bosons has $N+1$ equidistant energy levels 
$E_k=-J(N-2k)$ ($k=0,1,...,N$) with eigenstates 
\beq
\label{eigenstates0}
|E_k\rangle=\frac{2^{-N/2}}{\sqrt{k!(N-k)!}}
(a_l^\dagger+a_r^\dagger)^k(a_l^\dagger-a_r^\dagger)^{N-k}|0,0\rangle
\ ,
\eeq
where the normalization follows directly from $(a^\dagger)^l|0\rangle=\sqrt{l!}|l\rangle$.
Thus, the fastest oscillations occur with frequency $NJ$ and the characteristic parameter for the evolution
is $J\tau$.
The spectral weights $q_{n,k}:=\langle n,N-n|E_k\rangle$
are explicitly calculated in Eq. (\ref{sp_weight1}) of App. \ref{app:eigenstates}.
In particular, for the special cases $n=0$ and $n=N$ we have
\beq
q_{0,k}=2^{-N/2}\frac{\sqrt{N!}}{\sqrt{k!(N-k)!}}(-1)^{N-k}
\ ,\ \ \
q_{N,k}=2^{-N/2}\frac{\sqrt{N!}}{\sqrt{k!(N-k)!}}
.
\eeq
These specific expressions can be entered into the FDR/FDT amplitude $\phi_{m;n0} $ of Eq. (\ref{transition_amp3}), 
using $D_k=e^{-iE_k\tau}=e^{iJ(N-2k)\tau}$.
Moreover, $Q_n$, the diagonal matrix with elements $q_{n,k}$ for fixed $n$, is real here, which enables us to write
\beq
\label{transition_amp3a}
\phi_{m;n0} =Tr[D(D-Q_nEDQ_n)^{m-1}Q^*_0EQ_n]
=Tr[D^{1/2}T^{m-1}D^{1/2}Q^*_0EQ_n]
.
\eeq
This gives immediately the reduced density matrix of Eq. (\ref{FDR/FDT_relation}), the EE of Eq. (\ref{ent_entropy}) 
and the ES of Eq. (\ref{espectrum}).
Some examples for the eigenvalues of the monitored evolution matrix $T$ are presented in Fig. \ref{fig:ee0}.

\begin{figure}[t]
\begin{center}
\includegraphics[width=0.45\columnwidth]{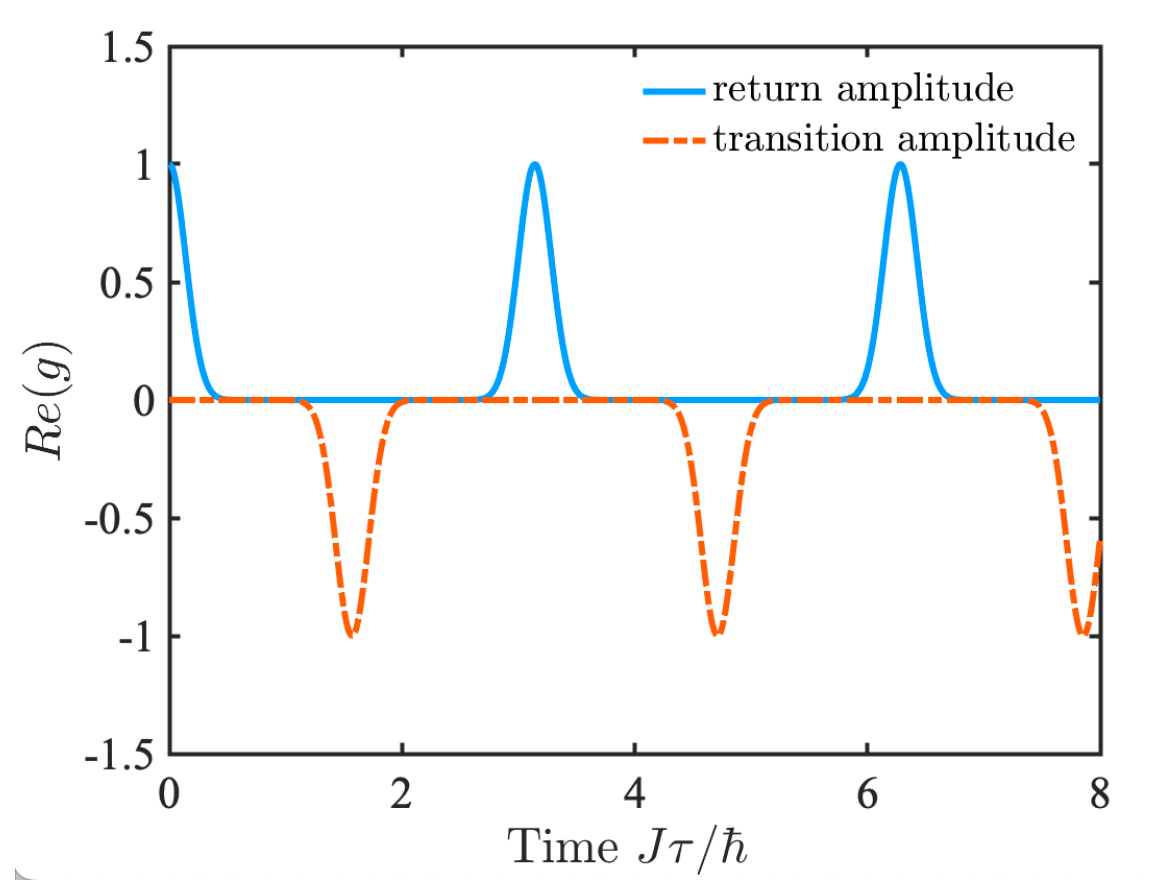}
\caption{
Unitary evolution:
Real parts of the return amplitude $u_1(\tau)$ 
($|0,N\rangle\to |0,N\rangle$) and the transition amplitude $v_1(\tau)$ 
($|0,N\rangle\to |N,0\rangle$) as a function of the dimensionless time $J\tau/\hbar$ for 50 non-interacting 
bosons.
}
\label{fig:dyn1}
\end{center}
\end{figure}

Returning to the 
unitary evolution, the amplitudes $u_m$ and $v_m$ of Eq. (\ref{matrix_elements}) are determined by a 
multiple of the frequency $J\tau$:
\beq
\label{non_int_exp}
u_m=\langle 0,N|e^{-iHm\tau}|0,N\rangle=\cos^N(mJ\tau) , \ \ \
v_m=\langle N,0|e^{-iHm\tau}|0,N\rangle=(-i)^N\sin^N(mJ\tau)
,
\eeq
which are periodic with $m\tau J/\hbar=2\pi$ or periodic with $m\tau J/\hbar=\pi$ for even $N$ 
(cf. Fig. \ref{fig:dyn1}). 
For a very short time (i.e., for $mJ \tau\ll 1/\sqrt{N}$) we have a Gaussian decay of the Fock state $|0,N\rangle$ as
\beq
\label{decay0}
u_m=\cos^N(mJ\tau)\sim e^{-Nm^2J^2\tau^2/2}
,
\eeq
as also illustrated in Fig. \ref{fig:dyn1}.
This non-exponential behavior in $\tau$ reflects the quantum Zeno effect \cite{misra1977zeno}.
Finally, the Fourier transformed  unitary amplitudes for $|\psi_0\rangle=|0,N\rangle$ and $|\psi\rangle=|N,0\rangle$ read
\beq
\label{g-u-ni}
{\hat u}(z)=2^{-N}\sum_{k=0}^N {{N}\choose{k}} \frac{z}{e^{iE_k\tau}-z}
\ ,\ \ \ 
{\hat v}(z)=(-2)^{-N}\sum_{k=0}^N {{N}\choose{k}} \frac{(-1)^k z}{e^{iE_k\tau}-z}
\ .
\eeq
They can be used to calculate the FDR/FDT amplitudes as the residue of the corresponding Cauchy integrals.

\section{Discussion of the results}
\label{sect:discussion}

The results of the previous section will now be used to calculate the FDR/FDT probabilities, the EE
and the ES for specific realizations of the model. To this end we note that the parameters of 
the bosonic system are the number of bosons $N$, the number of measurements $m-1$ and the time steps between
measurements $\tau$. The latter always appears in the combination with the tunneling rate $J$ as $J\tau$. This 
is a consequence of the fact that we have non-interacting bosons, where tunneling is the only mechanism of the evolution.
The combination $J\tau/\hbar$ provides a dimensionless time step in our system, which we will use subsequently.

\begin{figure}[t]
\includegraphics[width=0.9\columnwidth]{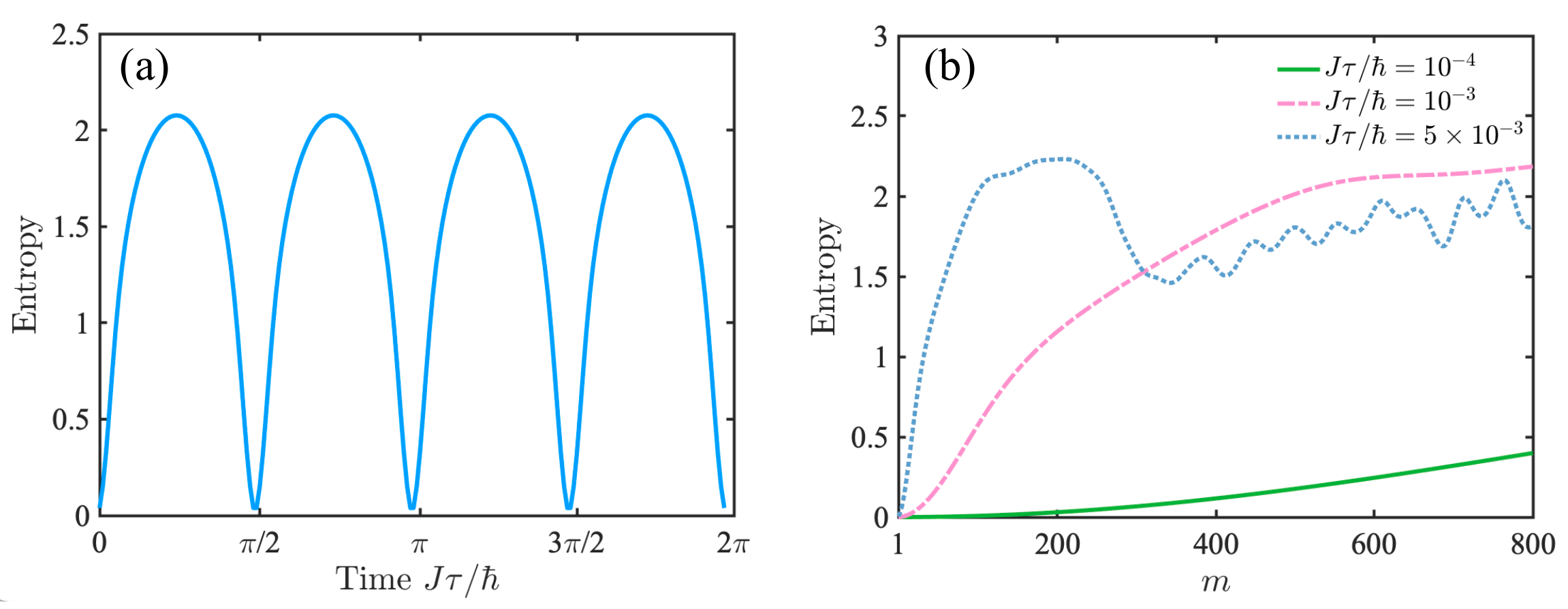} 
\caption{
(a) The unitary evolution of $N=20$ bosons for the time $J\tau/\hbar$ is periodic. 
(b) The monitored evolution of the entanglement entropy for a high frequency of measurements reflects the 
quantum Zeno effect, where an increasing measurement frequency reduces the entanglement entropy [green (lower) curve].
}
\label{fig:entanglement_entropy0}
\end{figure}

The characteristic features of the unitary dynamics defined in Eq. (\ref{non_int_exp}) is visualized in Fig. 
\ref{fig:dyn1}, which indicates a smooth variation of the amplitudes over time for the return to the initial state 
$|0,N\rangle\to |0,N\rangle$ and for the transition of all bosons to the other well $|0,N\rangle\to |N,0\rangle$. 
This periodic behavior is also reflected by the unitary evolution of the EE in Fig. \ref{fig:entanglement_entropy0}a,
which vanishes when all bosons are either in the left or in the right well. Repeated measurements will substantially
affect this periodic behavior.

\begin{figure}[h]
\psfrag{piJ}{$\pi J\tau/\hbar$}
\includegraphics[width=0.9\columnwidth]{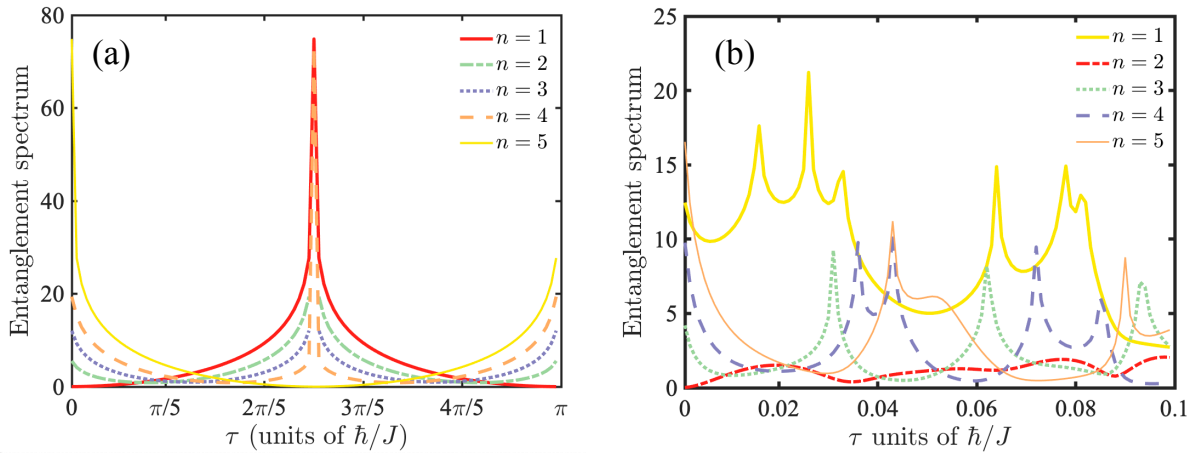}
\caption{
The entanglement spectrum for the 5 levels of $N=4$ bosons for $m=1$ (a) and $m=50$ (b) measurements 
as a function of the time step units $\pi J\tau/\hbar$ and $10^{-3}J\tau/\hbar$, respectively.
}
\label{fig:entanglement_spectrum1}
\end{figure}

\begin{figure}[h]
\includegraphics[width=0.85\columnwidth]{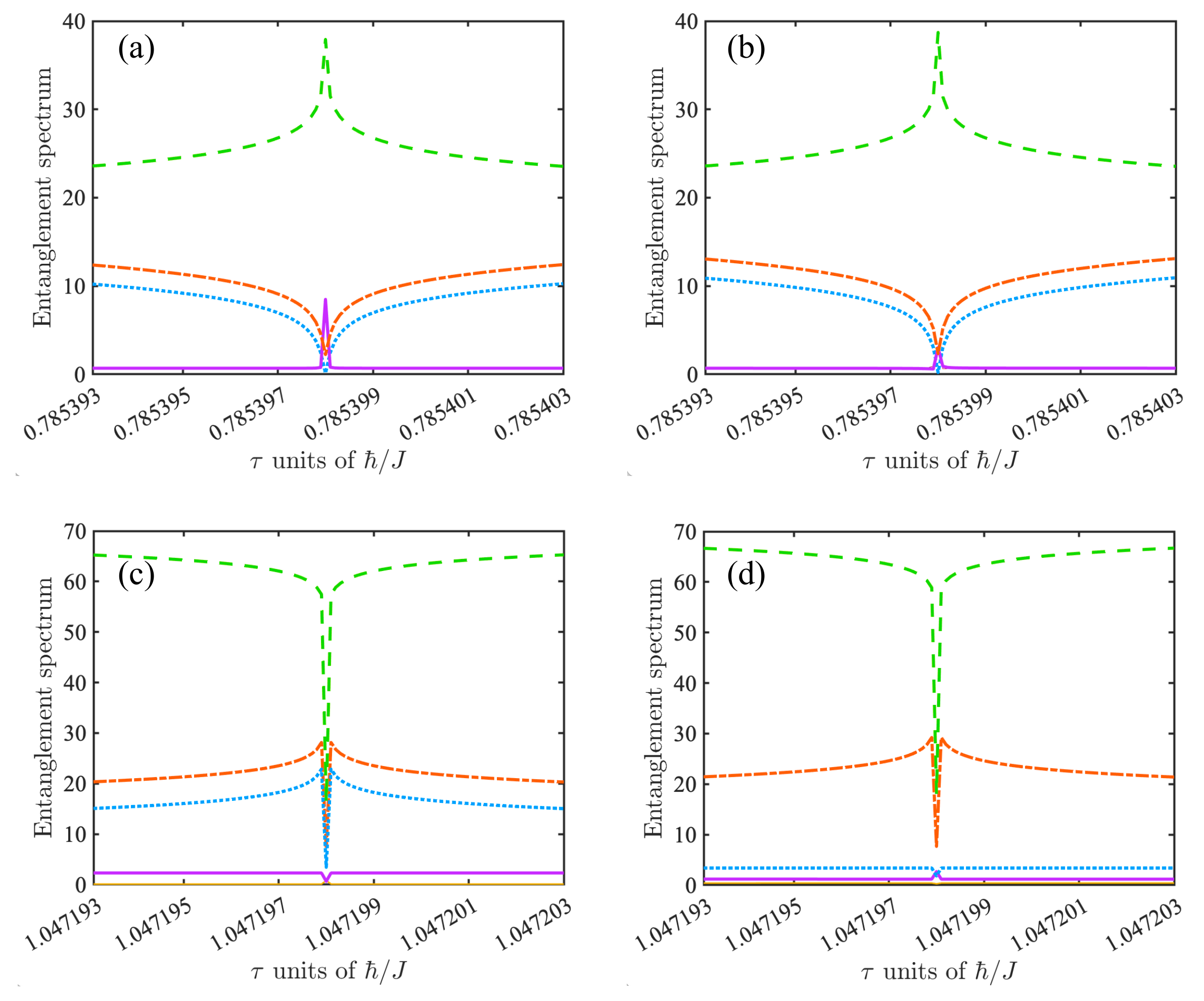}
\caption{
Entanglement spectrum for $N=4$ centered around $J\tau=\pi\hbar/4$ (a) and centered around $J\tau=\pi\hbar/3$ (b).
The left panels present $m=50$ measurements, the right panels $m=51$ measurements. The
jump of the green dashed level in (b) causes the
switching effect of the entanglement entropy between two values, as illustrated in Fig. \ref{fig:ee_4}(b).
}
\label{fig:eevses_4a}
\end{figure}

\begin{figure}
\includegraphics[width=0.9\columnwidth]{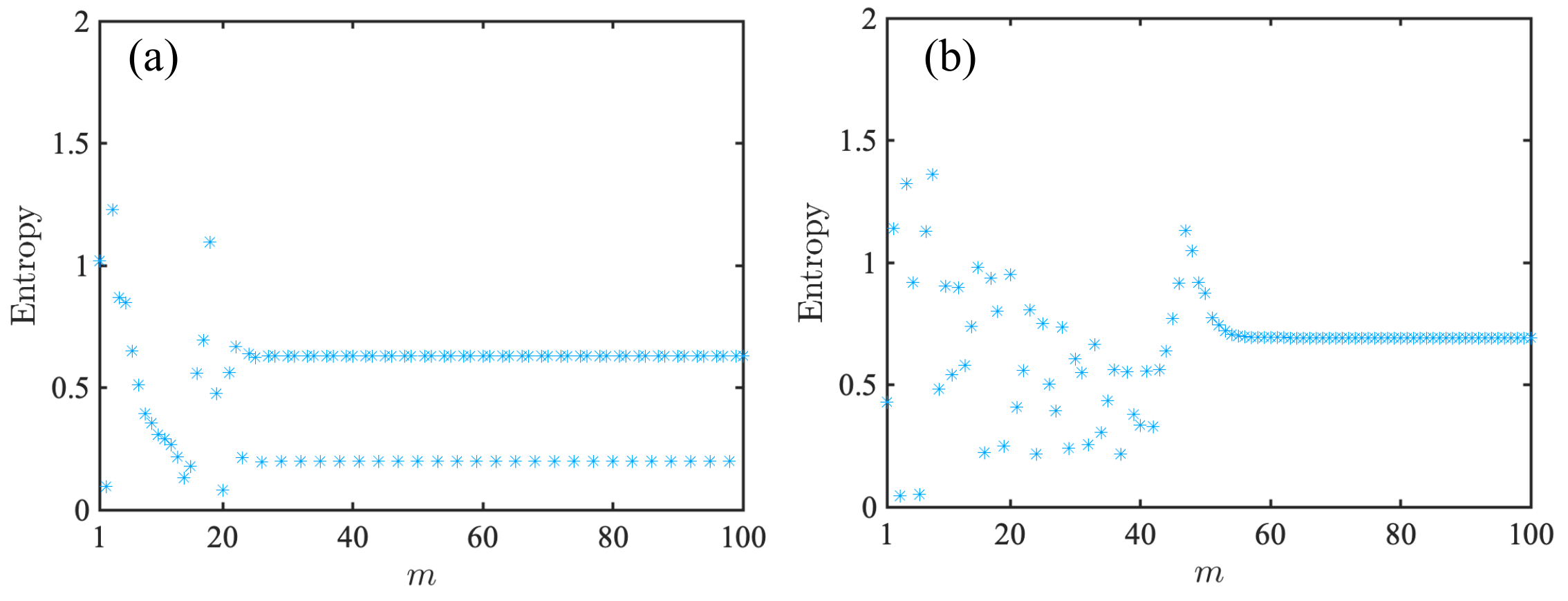}
\caption{
The entanglement entropy for $N=4$ is plotted for $J\tau=\pi\hbar/4$ (a) and $J\tau=\pi\hbar/3$ (b).
}
\label{fig:ee_4}
\end{figure}


Without measurement (i.e., for $m=1$) the EE is determined by the unitary amplitudes for all transitions 
$|0,N\rangle\to|n,N-n\rangle$, which are smooth and periodic in time
\beq
\phi_{1,n0}=e^{iNJ\tau}\sum_{k=0}^Ne^{-2ikJ\tau}q_{n,k}q_{0,k}
.
\eeq
This expression, together with Eq. (\ref{orthogonal1}), gives for $J\tau=0 \ ({\rm mod}\ 2\pi)$ and
$J\tau/\hbar=\pi \ ({\rm mod}\ 2\pi)$
\beq
\phi_{1,n0}=e^{iNJ\tau}\delta_{n0}
,
\eeq
implying ${\cal S}_2(\tau,N,1)=0$. This is reflected in the plot of ${\cal S}_2(\tau,N,1)$ of 
Fig. \ref{fig:entanglement_entropy0}a, which indicates also a vanishing EE at $J\tau/\hbar=\pi/2 \ ({\rm mod}\ 2\pi)$.
The periodicity does not depend on the number of bosons $N$, while the 
value of the EE increases with $N$. This is remarkable because the eigenvalues as well as the weight 
$q_{n,k}$ depend strongly on $N$.
The behavior of the EE is affected by measurements, depending on the time steps between the measurements.
As visualized in Fig. \ref{fig:entanglement_entropy0}b for $N=20$ bosons, for a very short time step $J\tau/\hbar$ between
measurements the periodic behavior of the unitary evolution disappears.
The corresponding ES in Fig. \ref{fig:entanglement_spectrum1} reveal that the level crossings 
are more complex in case of the monitored evolution and they take place on a much shorter time scale.

As already mentioned in the discussion of the FDR/FDT probabilities, the unitary evolution between measurements is
characterized by the phase factors $\exp(-iJ\tau m/\hbar)$, which is periodic for $m=l$ if $J\tau=2\pi\hbar/l$. In other
words, if the time step is a fraction of $\pi$, we expect a special behavior of the monitored evolution, which might
be reminiscent of the periodic behavior of the unitary evolution. But how does the monitored evolution depend on $l$?
This we will analyze for $N=4$ bosons, by comparing $l=3$ and $l=4$. 
Fig. \ref{fig:eevses_4a} the four plots visualize how the ES changes from $J\tau=\pi\hbar/l$ at the center to 
$(\pi/l\pm 5\cdot10^{-5})\hbar$ at the boundaries. The left panels represent the ES for $m=50$ measurements, 
the right panels the 
ES for $m=51$ measurements.  There are level degeneracies only at $\pi\hbar/l$, while in the narrow vicinity
the levels are well separated and the spectrum is symmetric with respect to $\pi\hbar/l$. It should be noticed though that
for $l=4$ (Fig. \ref{fig:eevses_4a}a)  the lowest level is two-fold degenerated, whereas for $l=3$ 
(Fig. \ref{fig:eevses_4a}b) there is not such a degeneracy. Another remarkable difference between $l=3$ and $l=4$
consists of the level change when the number of measurements changes from $m=50$ (left) to $m=51$ (right). While the
lowest two levels for $l=4$ are not affected by this change, there is a drastic change for $l=3$. The latter has two
low levels for $m=50$ but three low levels for $m=51$. The qualitative difference between these two $m$ values 
reflects the fact that for an odd $l$ only an odd $m$ can satisfy the condition $mJ\tau=2\pi\hbar$ for 
peridiodicity of the phase factor. To understand the effect of this $m$-dependence on the EE,
we use the definitions of the ES and the EE in Eqs. (\ref{ent_entropy}) and (\ref{espectrum}) and express the EE by
the levels of the ES as
\beq
{\cal S}_2=-\log_2\left(\sum_{n=0}^Ne^{-2\xi_{m;n}}\right)
,
\eeq
where the sum is reminiscent to the sum of Boltzmann weights in statistical mechanics. 
Therefore, only small values of $\xi_{m;n}$ (i.e., low levels of the ES) contribute substantially to the EE.
This means that the EE changes for $m\to m+1$ when $l=3$ but it remains unchanged for $l=4$. This is what we see
in Fig. \ref{fig:ee_4}. After some fluctuations for small values of $m$, the EE becomes stationary: For $l=4$ there is
just one stationary value (Fig. \ref{fig:ee_4}a), while for $l=3$ the EE switches between two stationary values
(Fig. \ref{fig:ee_4}b).

The above analysis relies on the fractional form $J\tau/\hbar=\pi/l$. The behavior of the EE for other values of the
time step between measurements can change drastically and may lead to a strongly fluctuating behavior of the EE.
In general,  repeated measurements have two major effects on the evolution: they destroy the periodicity (recurrence)  
and they lead to more level crossings in the ES, as illustrated in Fig. \ref{fig:entanglement_spectrum1}. The origin of these 
effects is that the non-interacting bosons are coupled repeatedly in time to the measurement apparatus, which provides a 
similar effect as a local boson-boson interaction, since the measurement is performed on the same quantum state at
different times. The situation can be compared with the unitary evolution of bosons with an interaction $U\ne 0$ in the
Bose-Hubbard Hamiltonian 
\beq
\label{Bose-Hubbard1}
H=-J(a_l^\dagger a_r + a_r^\dagger a_l)+\frac{U}{2}(n_l^2+n_r^2)
.
\eeq

We previously studied this system and found a similar behavior of a fluctuating EE and level
crossings in the ES~\cite{PhysRevA.107.012413}. The fluctuations were removed by averaging over time intervals, 
an approach we could also apply in the present case with $U=0$. For the monitored evolution we could also 
introduce random time steps between measurements and average over their 
distribution~\cite{Ziegler_2021,Das_2022,Das_2022a,acharya2023tightbinding,Kulkarni_2023}.
As a disadvantage of such a time averaging, though, we would not be able to detect the switching of the EE
in Fig. \ref{fig:ee_4}b.

\section{Conclusion}

We have studied the monitored evolution of $N$ non-interacting bosons which tunnel between two wells.
The monitoring is carried out by repeated projective measurements. 
The effect of these measurements is studied in terms of FDR/FDT probabilities to determine quantitatively the monitoring.
From the FDR/FDT probabilities we have derived the reduced density matrix for one well, the EE and the ES. This is 
based on the relation (\ref{FDR/FDT_relation}) and enabled us to evaluate the EE and the ES directly from the
FDR/FDT probabilities. It turns out that the EE is quite sensitive to a change of model parameters;
i.e., on the number of bosons and the time step between two measurements. 
The rather complex behavior of the EE indicates that a single quantity, such as the EE, is quite limited for the characterization
of the entanglement in the present system. More details are revealed by the ES of  Eq. (\ref{espectrum}). 
It enables us to identify the statistical weight of each transition $|0,N\rangle\to|n,N-n\rangle$ to the monitored 
evolution individually. 
A characteristic feature of the ES is  level crossing. Although it already appears in the unitary evolution 
(cf. Fig.  \ref{fig:entanglement_spectrum1}a), it becomes much more complex 
for the monitored evolution in Fig. \ref{fig:entanglement_spectrum1}b)
Except for the crossing points, there is always a unique lowest level,
representing the dominant transition $|0,N\rangle\to|n,N-n\rangle$. The excitation to higher levels is important as long
they are close to the lowest levels. This effect is important when the level $\xi_{m;n}$ changes quickly with $m$.
This can happen near the special values $J\tau=\pi\hbar/l$ ($l$ integer), as demonstrated in Figs. \ref{fig:eevses_4a}.

Although our approach was employed only to non-interacting bosons, it is directly applicable to interacting bosons
as well. 
For instance, we can consider the two-site Bose-Hubbard model with the Hamiltonian of Eq. (\ref{Bose-Hubbard1}).
New regimes might appear due to the competition of particle tunneling, particle-particle interaction and the
interaction with the measurement apparatus.
This is an ambitious project for the future, in which, among other aspects, the role of Hilbert-space localization should 
be addressed.

\medskip
\textbf{Acknowledgements} \par 
We are grateful to Eli Barkai for the useful discussions. This research is supported by the Israel Science Foundation 
through Grant No. 1614/21 (Q.L.).

\appendix

\section{Eigenstates}
\label{app:eigenstates}

With the rotated basis $\{a^{}_\pm,a^\dagger_\pm\}$:
\[
a^\dagger_\pm:=\frac{1}{2}(a^\dagger_l\pm a^\dagger_r)
\ , \ \ 
a^{}_\pm=\frac{1}{2}(a^{}_l\pm a^{}_r)
\]
we obtain for the tunneling operator $a^\dagger_la_r+a^\dagger_ra_l$
\beq
\label{diagonal_basis}
a^\dagger_la_r+a^\dagger_ra_l=\frac{1}{2}[(a^\dagger_l+a^\dagger_r)(a_l+a_r)-(a^\dagger_l-a^\dagger_r)(a_l-a_r)]
=\frac{1}{2}(a^\dagger_+a^{}_+ -a^\dagger_-a^{}_-)
,
\eeq
where $(a^\dagger_l\pm a^\dagger_r)(a_l\pm a_r)$ are number operators.
Then we can directly show that $a_+$ and $a_-$ and their Hermitean conjugate commute when $a_l$ and $a_r$ commute.
As a consequence, the eigenstate reads
\[
|E_k\rangle=\frac{2^{-N/2}}{\sqrt{k!(N-k)!}}(a^\dagger_+)^k(a^\dagger_-)^{N-k}|0,0\rangle
\]
and the application of the tunneling operator yields
\[
\frac{1}{2}(a^\dagger_+a^{}_+ -a^\dagger_-a^{}_-)|E_k\rangle
=(N-2k)|E_k\rangle
.
\]

The knowledge of the eigenstates enables us to calculate the spectral weights as scalar products
\beq
\label{sp_weight1}
q_{n,k}:=\langle n,N-n|E_k\rangle
=\frac{2^{-N/2}}{\sqrt{k!(N-k)!}}\langle n,N-n|(a_l^\dagger+a_r^\dagger)^k(a_l^\dagger-a_r^\dagger)^{N-k}
|0,0\rangle
\]
\[
=\frac{2^{-N/2}}{\sqrt{k!(N-k)!}}\langle n,N-n|\sum_{l=0}^k{k\choose l}(a^\dagger_l)^l(a^\dagger_r)^{k-l}
\sum_{l'=0}^{N-k}{N-k\choose l'}(a^\dagger_l)^{l'}(-a^\dagger_r)^{N-k-l'}|0,0\rangle
,
\]
and since the left and right operators commute, we obtain after reordering
\[
=\frac{2^{-N/2}}{\sqrt{k!(N-k)!}}\sum_{l=0}^k{k\choose l}\sum_{l'=0}^{N-k}{N-k\choose l'}(-1)^{N-k-l'}
\langle n,N-n|(a^\dagger_l)^{l+l'}(a^\dagger_r)^{N-l-l'}|0,0\rangle
.
\eeq
Due to the orthogonality of the states the sum vanishes unless $l+l'=n$. Then from the $l$ summation there
are two constraints for $l'$:
\[
0\le l'=n-l
\ ,\ \ \ 
n-l=l'\le N-k
,
\]
which are equivalent to $n+k-N\le l\le n$, such that
\[
\cases{
n+k-N\le l\le n & for $n+k-N>0$ \cr
0\le l\le n & for $n+k-N\le0$ \cr
}
.
\]
This result, together with  $(a^\dagger)^l|0\rangle=\sqrt{l!}|l\rangle$, implies for the scalar product
\beq
\label{spectral_coefficient}
q_{n,k}
=2^{-N/2}\sqrt{{N\choose k}\Big/{N\choose n}}\cases{
\sum_{l=0}^{\min\{k,n\}}{k\choose l}{N-k\choose n-l}(-1)^{N-k-n+l} & for $n+k\le N$ \cr
\sum_{l=n+k-N}^{\min\{k,n\}}{k\choose l}{N-k\choose n-l}(-1)^{N-k-n+l} & for $n+k> N$ \cr
}
.
\eeq
The orthonormal condition of the states $|n,N-n\rangle$ and the states $|E_k\rangle$, respectively, lead to
\beq
\label{orthogonal1}
\sum_{n=0}^Nq_{n,k}q_{n,k'}=\delta_{kk'}
\ \ \ {\rm and}\ \ \
\sum_{k=0}^Nq_{n,k}q_{n',k}=\delta_{nn'}
.
\eeq

\section{Expansion of the first return/transition amplitude}
\label{app:expansion}

For the FDR/FDT amplitude 
\beq
\label{new_def}
\phi_{m+1,1}:=\langle\psi|\left(e^{-i\tau H}\Pi\right)^{m}e^{-i\tau H}|\psi_0\rangle
=\langle\psi|e^{-i\tau H}\left(\Pi e^{-i\tau H}\right)^{m}|\psi_0\rangle
\ ,\ \ 
\Pi={\bf 1}-|\psi_0\rangle\langle\psi_0|
\ ,
\eeq
we obtain two equivalent recursion relations, namely
\beq
\phi_{m+1,1}=\phi_{m,2}'-u_1\phi_{m,1}
\ \ {\rm with}\ \ 
\phi_{m,k}'=\langle\psi|e^{-ik\tau H}\left(\Pi e^{-i\tau H}\right)^{m-1}|\psi_0\rangle
\ ,\ \ 
u_k=\langle\psi_0|e^{-iHk\tau}|\psi_0\rangle
\label{recurs1}
\eeq
from the third expression in Eq. (\ref{new_def}) and
\beq
\phi_{m+1,1}=\phi_{m,2}-\phi_{m,1}v_1
\ \ {\rm with}\ \ 
\phi_{m,k}=\langle\psi|\left(e^{-i\tau H}\Pi \right)^{m-1}e^{-ik\tau H}|\psi_0\rangle
\ ,\ \ 
v_k=\langle\psi|e^{-iHk\tau}|\psi_0\rangle
\label{recurs2}
\eeq
from the second expression in Eq. (\ref{new_def}). Then the iteration of Eq. (\ref{recurs1}) yields
with $\phi_{m}\equiv \phi_{m,1}$
\beq
\phi_{m}
=v_m-\sum_{j=1}^{m-1}u_{m-j}\phi_{j}
\ ,\ \ \
\phi_{1}=v_1
\label{app:1}
\eeq
and the iteration of Eq. (\ref{recurs2})
\beq
\label{app:2}
\phi_{m}=v_m-\sum_{j=1}^{m-1} \phi_{m-j}v_j
\ ,\ \ \
\phi_{1}=v_1
\ .
\eeq

Using the vector notation ${\vec\phi}:=(\phi_{1},\phi_{2},...,\phi_{m})$, 
${\vec v}=(v_1,v_2,...,v_m)$, with the matrix $\Gamma$ of Eq. (\ref{gamma_matrix})
and with the matrix ${\hat\phi}=(\phi_{m-j})$, we can write for Eqs. (\ref{app:1}) and (\ref{app:2}) 
\beq
({\bf 1}+\Gamma){\vec\phi}={\vec v}
\ ,\ \ 
{\vec\phi}=({\bf 1}-{\hat\phi}){\vec v}
,
\eeq
which yields $({\bf 1}-{\hat \phi})({\bf 1}+\Gamma)={\bf 1}$ and Eq. (\ref{final_eq}), respectively.




\end{document}